\documentclass[conference]{IEEEtran}
\usepackage{blindtext, graphicx, booktabs}
\usepackage{amsmath,amssymb}
\usepackage{cite} 

\IEEEoverridecommandlockouts                              

\title{\LARGE \bf
Optimization of Vehicle Connections in V2V-based Cooperative Localization
}

\begin{document}

\author{\IEEEauthorblockN{Macheng Shen}
\IEEEauthorblockA{University of Michigan\\Department of Naval Architecture\\and Marine Engineering\\
Ann Arbor, USA\\
Email: macshen@umich.edu}
\and
\IEEEauthorblockN{Jing Sun}
\IEEEauthorblockA{University of Michigan\\Department of Naval Architecture\\and Marine Engineering\\
Ann Arbor, USA\\
Email: jingsun@umich.edu}
\and
\IEEEauthorblockN{Ding Zhao}
\IEEEauthorblockA{University of Michigan\\
Department of Mechanical Engineering\\
Ann Arbor, USA\\
Corresponding author:\\
zhaoding@umich.edu}\\
}


\maketitle

\begin{abstract}

Cooperative map matching (CMM) uses the Global Navigation Satellite System (GNSS) positioning of a group of vehicles to improve the standalone localization accuracy. It has been shown to reduce GNSS error from several meters to sub-meter level by matching the biased GNSS positioning of four vehicles to a digital map with road constraints in our previous work. While further error reduction is expected by increasing the number of participating vehicles, fundamental questions on how the vehicle membership of the CMM affects the performance of the GNSS-based localization results need to be addressed to provide guidelines for design and optimization of the vehicle network. The quantitative relationship between the estimation error and the road constraints has to be systematically investigated to provide insights. In this work, a theoretical study is presented that aims at developing a framework for quantitatively evaluating effects of the road constraints on the CMM accuracy and for eventual optimization of the CMM network. More specifically, a closed-form expression of the CMM error in terms of the road angles and GNSS error is first derived based on a simple CMM rule. Then a Branch and Bound algorithm and a Cross Entropy method are developed to minimize this error by selecting the optimal group of vehicles under two different assumptions about the GNSS error variance.
%
%
\end{abstract}
\section{INTRODUCTION}
Low-cost Global Navigation Satellite Systems (GNSS) are used for most mobile applications, with localization error of several meters. The limited accuracy of the low-cost GNSS is mainly due to the atmospheric error, referred to as the common error, as well as receiver noise and multipath error, referred to as the non-common error. Various techniques have been used to reduce GNSS localization error such as precise point positioning (PPP) \cite{PPP}, real time kinematics (RTK) \cite{RTK} and sensor fusion \cite{fusion}. Nonetheless, improving the localization accuracy of the widespread GNSS without incurring additional hardware and infrastructure costs has motivated recent research activities on cooperative GNSS localization \cite{CL_GNSS} and cooperative map matching (CMM) \cite{rohani2016novel}, \cite{shen2017improving}, \cite{shen2016enhancement}. \\
\indent CMM improves the GNSS localization accuracy by matching the GNSS positioning of a group of connected vehicles to a digital map.  With the reasonable assumption that vehicles should travel on the roads, the road constraints can be applied to correct the GNSS positioning. CMM is much more powerful in exploiting the road constraints than ego-localization with map matching.
\begin{figure}[htbp]
  \centering
  \includegraphics[width=0.75\columnwidth]{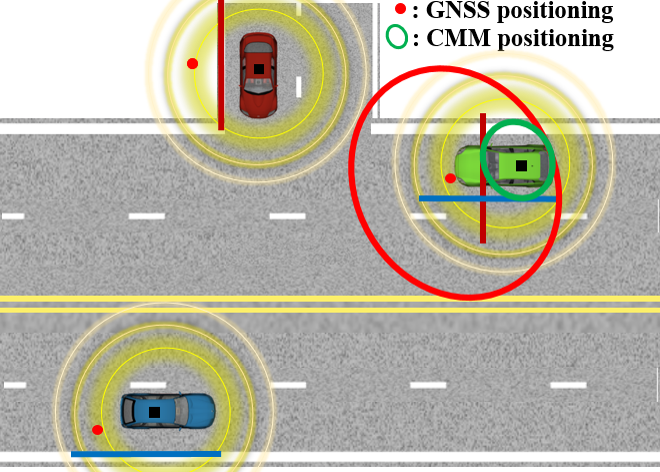}   
  \caption{Illustration of the positioning correction through CMM}
  \label{illustration}
\end{figure}

\indent Fig. \ref{illustration} illustrates how error-corrupted GNSS information is corrected through a naive CMM approach. The GNSS solutions, denoted as red dots, of all the three vehicles are biased away from their true positions, while the biases are highly correlated. Virtual constraints from the red and the blue vehicle are applied to restrict the positioning of the green vehicle. The underlying assumption for the application of the virtual constraints is that the GNSS positioning biases of the three vehicles are the same. This example manifests the effect of road constraints on the CMM localization. \\  
\indent The limited bandwidth of the Dedicated Short Range Communication (DSRC)-based connected vehicle network imposes an upper limit on the maximal number of connected vehicles to avoid frequent package loss  \cite{ramachandran2007experimental}, but typically the number of vehicles within the communication range is more than the upper limit. This limitation motivates the development of an optimization scheme that selects $M$ vehicles with optimal road constraints that minimizes CMM localization error, where $M$ is the maximal number of connected vehicles. To achieve this goal, it is necessary to quantify the effect of road constraints on the CMM localization error. This is, nevertheless, not easily achievable as the algorithms that implement CMM are usually sophisticated, and there is no analytic expression relating the localization error with the road constraints for these algorithms.\\ 
%
%
\indent Recently, two different CMM algorithms have been developed for GNSS common error estimation problem, i.e., a non-Bayesian particle-based approach in Rohani et. al. \cite{rohani2016novel} and a Bayesian approach based on a Rao-Blackwellized Particle Filter in our previous work \cite{shen2017improving}, \cite{shen2016enhancement}. One common feature of these two CMM algorithms is that the probabilistic property of the GNSS error is utilized. This increases the localization accuracy and robustness, but also leads to complicated mapping between the localization error and the road constraints, which makes the analysis of the effects of road constraints rather difficult.\\  
\indent In our previous work \cite{shen2016impact}, the correlation between the localization error and the road constraints is quantified analytically. More specifically, a closed-form expression of the estimated localization error in terms of the road angles as well as the GNSS error is derived based on a simple CMM rule that neglects the probabilistic property of the GNSS error as well as historical GNSS measurements. As a result of these simplifications, this closed-form expression is not sufficient to accurately predict the localization error of the particle filter based CMM algorithm. Nonetheless, it can be used to select the optimal set of connected vehicles, on which the CMM error is expected to be minimized with the particle filter based CMM algorithm. In this work, two algorithms are developed to optimize the connected vehicle selection such that the closed-form expression of the estimated localization error is minimized. These results provide a guideline for the implementation of CMM.\\
\indent In the following sections, details about the derivation of the CMM localization error and the optimization algorithms are presented. In Section 2, the notions in CMM are introduced. Then a simple CMM rule is applied to derive an analytic expression of the localization error. Based on this error formula, in Section 3, two algorithms are presented to minimize the localization error through the optimal selection of connected vehicles. A Branch and Bound (B\&B) \cite{clausen1999branch} searching algorithm is developed in the case that all the vehicles have the same non-common GNSS error variance. A Cross Entropy (CE) method \cite{botev2013cross} combined with a heuristic pre-selection step is developed in the case that vehicles have different non-common GNSS error variances. In Section 4, the contributions and conclusions are summarized.
\section{CMM localization error}
In this section, we propose a framework of vehicle positioning within a reference road framework to facilitate the analytic investigation.\\
\indent The following assumptions are essential in the derivation:
\begin{enumerate}
 \item The GNSS common error is the same for all the connected vehicles within the vehicular network. 
 \item The road side can be locally approximated as a straight line.
 \item The GNSS non-common error is small enough such that the exact expression for the estimation error can be approximated by its first order linearization with respect to the non-common error.
 \item The GNSS non-common error is a random variable that obeys the Gaussian distribution.\\
 \end{enumerate}
 \indent The first assumption is reasonably accurate as long as the connected vehicles are geographically close to each other, for example, within several miles. The second and the third assumptions are made for mathematical convenience. If they are violated, however, the exact expression of the estimation error will still be valid but the asymptotic approximation will be inaccurate. The last assumption has been experimentally verified in \cite{shen2017improving} under open sky conditions.
 
\begin{figure}[htbp]
  \centering
  \includegraphics[width=0.8\columnwidth]{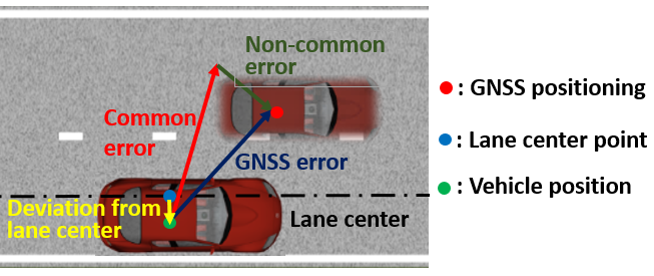}   
  \caption{Illustration of the notations used in CMM}
  \label{notation}
\end{figure}
\indent Consider a network of connected vehicles, the coordinate of GNSS positioning of the $i$-th vehicle can be decomposed into a superposition of the coordinate of a point on the corresponding lane center, the deviation of the vehicle from the lane center, the common error and the non-common error as illustrated in Fig. \ref{notation}. The grayscale image of the vehicle represents the GNSS positioning. This decomposition can be expressed mathematically as
\begin{equation}
x^G_i=x^L_i+x^D_i+x^C+x^N_i, i=1,2,...,N,
\end{equation}
where $x^G_i$ is the GNSS positioning of the $i$-th vehicle, $x^L_i$ is the closest point on the center of the lane from the vehicle, $x^D_i$ is the deviation of the vehicle coordinates from $x^L_i$, $x^C$ is the GNSS common error, $x^N_i$ is the GNSS non-common error.\\
\indent The fact that all the vehicles travel on the roads can be expressed as a set of inequalities
\begin{equation}
g_i(x^G_i-x^C-x^N_i)<0.
\label{eq2}
\end{equation}
\indent Applying the second assumption, the constraint functions $g_i$ have simple analytic forms
\begin{equation}
g_i(x)=(x-x^L_i)\cdot n_i-w,
\end{equation}
where $\{\cdot\}$ is the dot product operator, $n_i$ is the unit vector normal to the lane center point towards outside of the road and $w$ is the half width of the lane. \\
\indent Eq. (\ref{eq2}) can be interpreted as the feasible set of the common error given the GNSS positioning and the non-common error. The non-common error is unknown, however, to the implementation of CMM. Thus, the following approximation of the feasible set by neglecting the non-common error is used instead of the exact feasible set,
\begin{equation}
\begin{aligned}
\Omega &=\{\tau|\bigcap\limits_{i=1}^{N}g_i(x^G_i-\tau)<0\}\\
&=\{\tau|\bigcap\limits_{i=1}^{N}g_i(x^L_i+x^C+\tilde{x}^N_i-\tau)<0\}\\
&=\{\tau|\bigcap\limits_{i=1}^{N}\tilde{g}_i(x^C+\tilde{x}^N_i-\tau)<0\},
\end{aligned}
\end{equation}
where 
\begin{equation}
\tilde{x}^N_i \triangleq x^D_i+x^N_i
\end{equation}
and 
\begin{equation}
\tilde{g}_i(x)\triangleq g_i(x+x^L_i)=x\cdot n_i-w.
\end{equation}
A point estimator of the common error is taken as the average over the approximate feasible set $\Omega$,
\begin{equation}
\hat{x}^C=\frac{1}{S}\int\limits_\Omega \tau dA, S=\int\limits_\Omega dA,
\end{equation}
where $\tau$ is the dummy integration variable and $dA$ is the area element.\\
\indent The estimation error, that is the difference between the true common error and the estimated common error, is of practical interest, which can be evaluated,
\begin{equation}
\begin{aligned}
e&=x^C-\hat{x}^C=x^C-\frac{1}{S}\int\limits_\Omega \tau dA\\
&=\frac{1}{S}\int\limits_\Omega (x^C-\tau) dA=\frac{1}{S}\int\limits_{\Omega'} \tau' dA,
\end{aligned}
\label{eq8}
\end{equation}
where 
\begin{equation}
\tau'=x^C-\tau,
\end{equation}
and
\begin{equation}
\Omega'=\{\tau'|\bigcap\limits_{i=1}^{N}\tilde{g}_i(\tilde{x}^N_i+\tau')<0\}.
\label{eq10}
\end{equation}
\indent Eq. (\ref{eq8}) and (\ref{eq10}) states that the estimation error equals to the geometric center of the intersection of the road constraints perturbed by the composite non-common error $\tilde{x}^N_i$. \\
\indent Eq. (\ref{eq8}) is a random variable as a nonlinear function of the non-common error. The third assumption implies that (\ref{eq8}) can be linearized with respect to the non-common error:
\begin{equation}
e=e_0+\Delta e=e_0+\frac{C\tilde{X}}{S_0},
\label{eq11}
\end{equation}
where 
\begin{equation}
e_0=\frac{1}{S_0}\int\limits_{\Omega_0} \tau' dA,
\end{equation}
\begin{equation}
\Omega_0=\{\tau'|\bigcap\limits_{i=1}^{N}\tilde{g}_i(\tau')<0\},
\end{equation}
\begin{equation}
\tilde{X}=[\tilde{x}^N_1\cdot n_1,\tilde{x}^N_2\cdot n_2,...,\tilde{x}^N_N\cdot n_N]^T,
\end{equation}
and 
\begin{equation}
C=S_0\frac{\partial e}{\partial \tilde{X}}.
\end{equation}
$C$ is a $2\times N$ matrix whose components are related to the geometric quantities of the road constraints.\\
\indent The condition under which the linearization (\ref{eq11}) is valid is
\begin{equation}
||\tilde{X}||_\infty\ll\frac{2\pi w}{N},
\end{equation}
where $w$ is the half width of the lane.\\
\indent With the fourth assumption that each non-common error obeys independent Gaussian distribution with zero mean, i.e., $\tilde{X}\sim \mathcal{N}(0_{N\times 1}, diag(\sigma^2_1,\sigma^2_2,...,\sigma^2_N))$, the expectation of the square error is
\begin{equation}
E_X[e^2]=e^2_0+\frac{1}{S^2_0}tr(L^TC^TCL),
\label{eq17}
\end{equation}
where $L=diag(\sigma_1,\sigma_2,...,\sigma_N)$ is the Cholesky decomposition of the joint covariance matrix.\\
\section{Optimizing the selection of vehicles for minimal localization error}
In practice, the maximal number of connected vehicles implementing CMM is limited by the finite communication bandwidth. One interesting problem of practical importance is to optimally select a group of $M(M \le N)$ vehicles out of the total $N$ available connected vehicles to implement CMM. This optimization problem can be stated as determining the indices $j_1,j_2,...,j_M$ such that the corresponding road angles $\theta_{j_1},\theta_{j_2},...,\theta_{j_M}$ and the non-common error variances $\sigma^2_{j_1},\sigma^2_{j_2},...,\sigma^2_{j_M}$ minimize the mean square error (\ref{eq17}) as an objective function.\\
\indent This optimization problem is combinatorial. A B\&B searching algorithm is developed to efficiently find the optimal solution when all the vehicles have the same non-common error variance. When the vehicles have different non-common error variances, it will be proved that the part of the localization error that does not depend on the non-common error would be minimized if the road angles obey a uniform distribution. Motivated by this optimality of the uniform distribution, a CE method with a heuristic pre-selection step is developed to find a near optimal solution efficiently. The performances of those optimization algorithms are illustrated on synthetic data.
\subsection{The B\&B algorithm}
In the case that all the non-common error variances have the same value, i.e., $\sigma^2_1=\sigma^2_2=...=\sigma^2_N=\sigma^2$, the objective function (\ref{eq17}) is a function of the road angles only. It is straightforward to verify that the corresponding continuous optimization problem has one global minimizer
\begin{equation}
\theta^T_{opt}=[0,\frac{2\pi}{M},\frac{4\pi}{M},...,\frac{2\pi(M-1)}{M}].
\end{equation}
\indent It should be noted that this is not the solution to the connected vehicle selection problem as the actual road angles cannot happen to be the components of $\theta^T_{opt}$. Nevertheless, this global minimizer can be used to derive the bound function which is an indispensable part of the B\&B method, described as follows.\\
\indent The objective function (17) can be approximated by its truncated Taylor expansion as 
\begin{equation}
J(\theta)=E_X[e^2]\approx J_0+\frac{1}{2}\Delta\theta^T H\Delta\theta,
\label{eq19}
\end{equation} 
where $J_0=J(\theta_{opt})$ is the minimum of the continuous problem, $H$ is the Hessian at the minimum and $\Delta\theta$ is the deviation from $\theta_{opt}$.\\
\indent The B\&B algorithm solves the combinatorial optimization problem by searching the solution space represented as an enumeration tree. A bound function is used to estimate the lower bound of the objective function values of the subtree rooted at the active node and prune the branches that are guaranteed not to lead to the minimum.\\
\indent The bound function for the vehicle selection problem is obtained by minimizing the continuous relaxation of (\ref{eq19}) subject to equality constraints, where the equality constraints come from all the ancestors of the currently active node. This constrained minimization is equivalent to a non-constrained minimization of a quadratic function defined on $\mathbb{R}^{M^*}$, where $M^*<M$, which can be efficiently solved.\\
\begin{figure}[htbp]
  \centering
  \includegraphics[width=1\columnwidth]{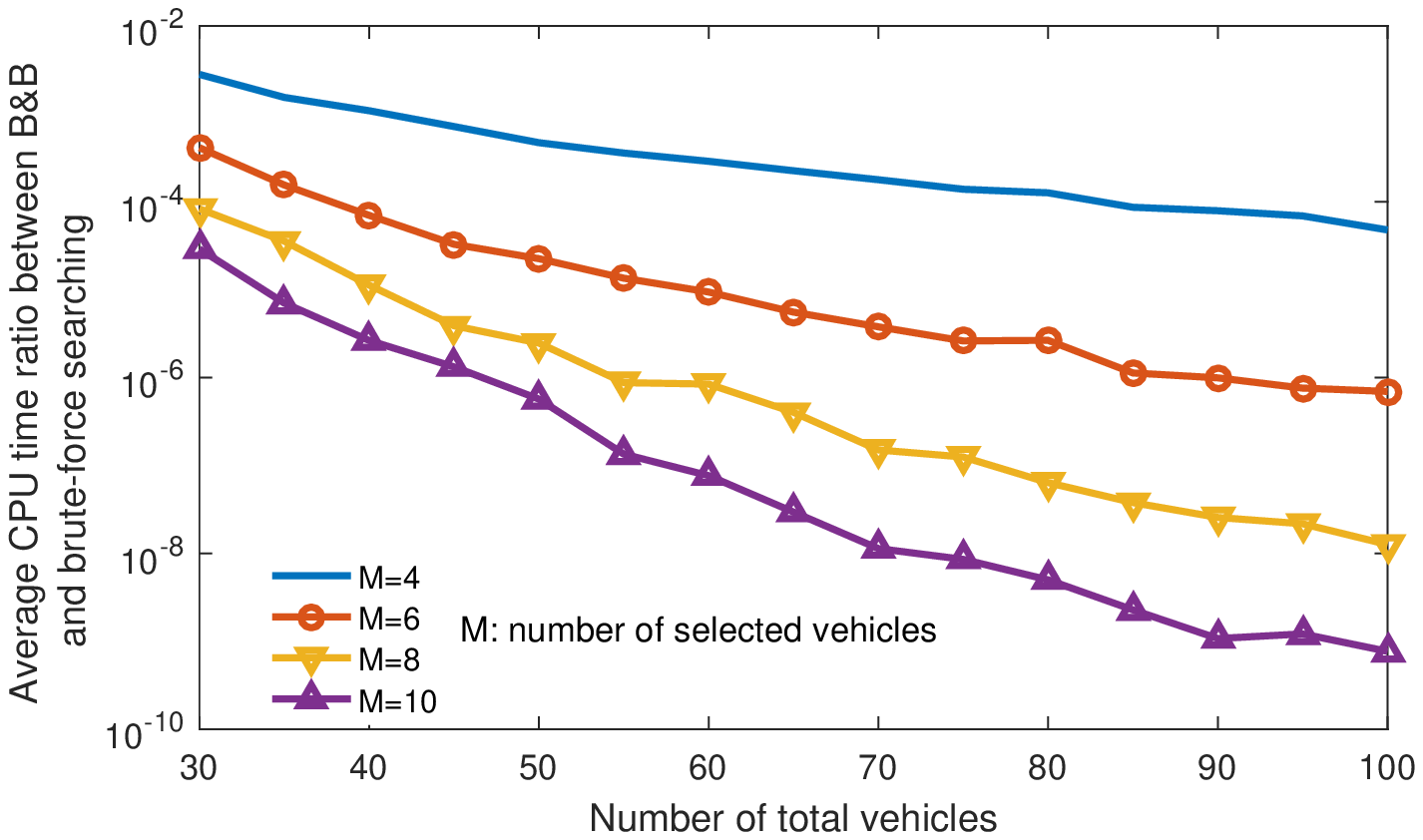}   
  \caption{Ratio between the average CPU time of the B\&B algorithm and that of the brute-force searching}
  \label{B&B}
\end{figure}
\indent As the B\&B method is guaranteed to find the optimal solution. The performance of this method is evaluated by the computational complexity compared with that of the brute-force searching. Fig. \ref{B&B} shows the ratio between the average CPU time of the B\&B method over 100 simulations and that of the brute-force searching. Instead of actually running the simulation for the brute-force searching, which can be computationally prohibitive, the computation time is estimated by the product of the average computation time for each objective function evaluation and the required number of evaluation. \\
\indent As the number of feasible solutions increases dramatically with the increase of $N$ and $M$, the B\&B method saves more computation time. For $N=100$ and $M=10$, the number of feasible solutions is $C^{10}_{100}=1.73\times 10^{13}$, while the average computation time using the B\&B method is only 0.5 s on MATLAB 2016a with an Intel i-7 6500U processor. Instead of evaluating the objective function for every possible combination, the B\&B method only did $O(10^4)$ evaluations on average, owing to the use of the bound function.
\subsection{Optimality of the uniform distribution}
In this section, a related problem will be considered, which serves as the foundation of a CE algorithm that solves the vehicle selection problem when the non-common error variances are different. Consider the road angles as random variables drawn from some distribution, the square error (\ref{eq17}) becomes a random variable as a function of the road angles as well as the non-common error variances. It will be proved that in the limit that the number of vehicles goes to infinity, the expectation of the geometric error $e^2_0$ is minimized if the angles of the roads on which the vehicles travel obey a uniform distribution. \\
\indent Considering an arbitrary continuous distribution of the road angle $p(\theta),\theta\in[0,2\pi)$, the periodic condition should be satisfied,
\begin{equation}
p(0)=p(2\pi^-),
\end{equation}
as $\theta=0$ and $\theta=2\pi$ represent the same angle.\\
\indent This periodicity motivates the following Fourier series expansion,
\begin{equation}
p(\theta)=\frac{1}{2\pi}+\sum_{m\in Z^*}C_mexp(im\theta),
\end{equation}
with
\begin{equation}
C_m=C^*_{-m},
\end{equation}
where the integer $m$ is the summation index, the asterisk denotes the complex conjugate and $i=\sqrt{-1}$ is the imaginary unit. The constant $\frac{1}{2\pi}$ ensures that the normalization condition is satisfied.\\
\indent In the limit that $N \xrightarrow{} \infty$, the leading order term of the localization error $e^2_0$ due to deviation of the geometric center can be approximated as
\begin{equation}
e^2_0=\frac{4w^2}{9}\frac{\sum_{i=1}^N \tan^2(\frac{\tilde{\theta_i}}{2})}{\pi^2} \approx \frac{w^2}{9}\frac{\sum_{i=1}^N \tilde{\theta_i}^2}{\pi^2},
\label{eq23}
\end{equation}
where $\tilde{\theta_i}$ is the difference between two adjacent angles $\theta_{i+1}$ and $\theta_i$
\begin{equation}
\tilde{\theta_i}=\begin{cases}
\theta_{i+1}-\theta_i & \text{for }i=1,2,...,N-1\\
\theta_1-\theta_N+2\pi & \text{for }i=N\\
\end{cases}
\end{equation}
\indent In order to derive the expectation of $e^2_0$, the distribution of $\tilde{\theta_i}$ denoted as $f(\tilde{\theta_i};N,p(\theta_i))$ will be considered first. $f(\tilde{\theta_i};N,p(\theta_i))$ is a nearest neighbor distribution, which satisfies the integral equation,
\begin{equation}
f(\tilde{\theta_i})=2Np(\theta_i)(1-\int^{\tilde{\theta_i}}_0f(\tau)d\tau),
\label{eq25}
\end{equation}
where the dependence on the parameters $N$ and $p(\theta_i)$ will be omitted hereafter.\\
\indent Together with the normalization condition, the solution to (\ref{eq25}) is
\begin{equation}
f(\tilde{\theta_i})=2Np(\theta_i)exp(-2Np(\theta_i)\tilde{\theta_i}).
\end{equation}
\indent The number of vehicles $N$ and the local density of the road angles $p(\theta_i)$ appear as parameters in the distribution. As the product $Np(\theta_i)$ increases, the angles distributed around $\theta_i$ become dense, thus increasing the probability of small differential angle $\tilde{\theta_i}$. The expectation of $\tilde{\theta_i}^2$ is
\begin{equation}
E[\tilde{\theta_i}^2]=\int^\infty_0 \tilde{\theta_i}^2 f(\tilde{\theta_i}) d\tilde{\theta_i}=\frac{1}{2N^2p^2(\theta_i)}.
\label{eq27}
\end{equation}
\indent Combining (\ref{eq23}) and (\ref{eq27}), the expectation of $e^2_0$ can be derived as follows,
\begin{equation}
E_{\theta}[e^2_0]=\frac{w^2}{9\pi^2}\sum ^N_{i=1} E[\tilde{\theta_i}^2]=\frac{w^2}{36N\pi^2}\sum^N_{i=1}\frac{1}{p^2(\theta_i)}\frac{2\pi}{N}.
\label{eq28}
\end{equation}
\indent The summation in (\ref{eq28}) can be interpreted as an integration as the number of the vehicles $N$ goes to infinity,
\begin{equation}
\begin{aligned}
\lim\limits_{N \to \infty }{\sum^N_{i=1}\frac{1}{p^2(\theta_i)}\frac{2\pi}{N}}&=\lim\limits_{N \to \infty }{\sum^N_{i=1}\frac{1}{p^2(\theta_i)}\Delta \theta}\\
&=\int^{2\pi}_{0}\frac{1}{p^2(\theta)}d\theta.
\end{aligned}
\label{eq29}
\end{equation}
\indent The Fourier expansion of $\frac{1}{p^2(\theta)}$ can be obtained and expressed in terms of the Fourier coefficients of $p(\theta)$, assuming the deviation from the uniform distribution is infinitesimal,
\begin{equation}
\begin{aligned}
\frac{1}{p^2(\theta)}&=4\pi^2-8\pi^3\sum_{m\in Z^*}C_mexp(im\theta)\\
&+16\pi^4(\sum_{m\in Z^*}C_mexp(im\theta))^2+O(C^3_m).
\end{aligned}
\label{eq30}
\end{equation}
\indent Substituting (\ref{eq30}) into (\ref{eq29}),
\begin{equation}
\lim\limits_{N \to \infty }{\sum^N_{i=1}\frac{1}{p^2(\theta_i)}\frac{2\pi}{N}}=8\pi^3+32\pi^5 \sum^{\infty}_{m=1}|C_m|^2.
\label{eq31}
\end{equation}
\indent Substituting (\ref{eq31}) into (\ref{eq28}),
\begin{equation}
E_{\theta}[e^2_0]\sim \frac{w^2}{36N\pi^2}(8\pi^3+32\pi^5 \sum^{\infty}_{m=1}|C_m|^2).
\end{equation}
\indent By taking $C_m=0, m\in Z^+$, which corresponds to the uniform distribution, the expectation of the square error $E_{\theta}[e^2_0]$ is minimized.\\
\subsection{The CE method based two-step algorithm}
The two-step algorithm is comprised of a heuristic pre-selection step to downsize the eligible vehicle population and a sampling-based searching step that applies the CE method for the optimization inspired by the existence of the optimal distribution, described as follows:
\subsubsection{Step one: pre-selection}
An important observation that motivates this pre-selection is that the objective function (\ref{eq17}) is monotonically increasing with respect to the non-common error variances $\sigma^2_i,i=1,2,...,M$. If two vehicles have close road angles, i.e., $\theta_i \approx \theta_j$ and large difference of the non-common error variance, e.g., $\sigma^2_i \ll \sigma^2_j$, then it is very unlikely that selecting the $j$-th vehicle will lead to the optimal solution as substituting the $j$-th vehicle with the $i$-th vehicle would probably make the objective function smaller. The following procedure is used to eliminate those unpromising vehicles:
\begin{itemize}
\item For all pairs that satisfy $\sigma^2_i < \sigma^2_j$:\\
Sample randomly $K$ pairs of groups of vehicles with indices: $[n_{l,1},n_{l,2}...,n_{l,M-1},i],[n_{l,1},n_{l,2}...,n_{l,M-1},j]$, where $l=1,2,...,K$, and evaluate the corresponding pairs of the objective function values $J_{l,i},J_{l,j}$.
\item If $J_{l,i} < J_{l,j}$ for all $l=1,2,...,K$:\\
Then eliminate the j-th vehicle from the selection candidates.
\end{itemize}
\subsubsection{Step two: CE method}
After the pre-selection step, a CE method is applied, motivated by the existence of the optimal distribution (the uniform distribution) proved in the previous section. The CE method searches for the minimizer of the objective function by iteratively sampling from a parameterized distribution and updating the parameter according to the samples with good performance. In this particular optimization problem, a $M$-dimensional Gaussian distribution of the road angle is used, which is described by the following pseudo-code:
\begin{itemize}
\item Iterate until the convergence criterion (for the distribution parameters) is satisfied:
\item Generate $k$ samples from the Gaussian distribution denoted as $[s_1,s_2,...,s_k]$, where each $s_i=[\theta_{i,1},\theta_{i,2},...,\theta_{i,M}],i=1,2,...,k$ is an $M$ dimension vector of angles that are not necessarily equal to any of the actual road angles.
\item Round the angles to the nearest road angles, which leads to the rounded samples $\tilde{s}_1,\tilde{s}_2,...,\tilde{s}_k$, and evaluate the values of the objective function denoted as $J_1,J_2,...,J_k$.
\item Update the mean and covariance parameter as $\mu=\frac{\sum^{k}_{i=1} I[J_i \geq \hat{\gamma}]\tilde{s}_i}{k_\rho}$ and $cov=\frac{\sum^{k}_{i=1} I[J_i \geq \hat{\gamma}](\tilde{s}_i-\mu)(\tilde{s}_i-\mu)^T}{k_\rho-1}$ \cite{botev2013cross}, where $\hat{\gamma}$ is the sample $(1-\rho)$- quantile of performance, $k_\rho$ is the corresponding number of samples and $I$ is the indicator function.
\end{itemize}
\indent The performance of this two-step algorithm is demonstrated on synthetic data with $N=50$ and $M=5$. In addition, the road angles are drawn from a uniform distribution and the non-common error variances are generated by $\sigma^2_i=(0.5+|v_i|)\mbox{ }m^2$, where $v_i\sim \mathcal{N}(0,1)$. The other parameters are: $K=10$, $k=1000$, $\rho=0.05$. The initial parameters of the Gaussian distribution are $\mu_0=[0,\frac{2\pi}{M},\frac{4\pi}{M},...,\frac{2(M-1)\pi}{M}]$ and $cov_0=diag([100(\frac{\pi}{M})^2,100(\frac{\pi}{M})^2...,100(\frac{\pi}{M})^2]^T_{M\times 1})$.

\begin{figure}[htbp]
  \centering
  \includegraphics[width=0.9\columnwidth]{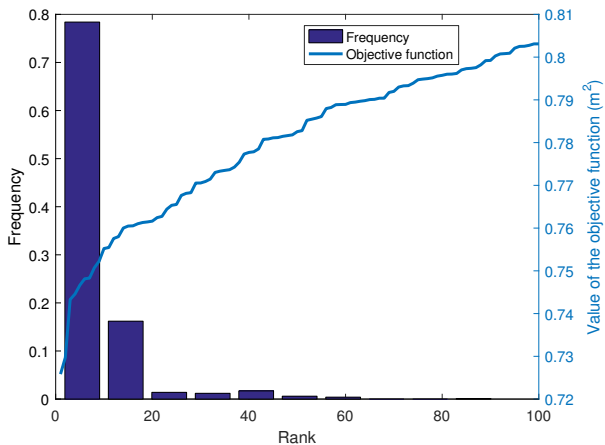}   
  \caption{Performance of the optimization algorithm; the total number of feasible solution is $C^5_{50}=2,118,760$.}
  \label{CE}
\end{figure}

\indent Fig. \ref{CE} shows the distribution of the optimization results sorted into increasing order of the objective function values based on 1000 simulations and the sorted values of the objective function corresponding to the best 100 combinations of $M$ vehicles (blue line). The total number of feasible solution is $C^5_{50}=2,118,760$, while the solutions found by the algorithm are among the best 20 ones in 95\% of the cases and all the results are within the best 100 ones. The average computation time is 1.6 s for each simulation.\\
\indent The pre-selection step plays an important role in achieving this performance. Without the pre-selection, the CE algorithm would waste a lot of searching on the samples that are unpromising to be optimal. As a result, the performance would degrade.

\begin{figure}[htbp]
  \centering
  \includegraphics[width=0.9\columnwidth]{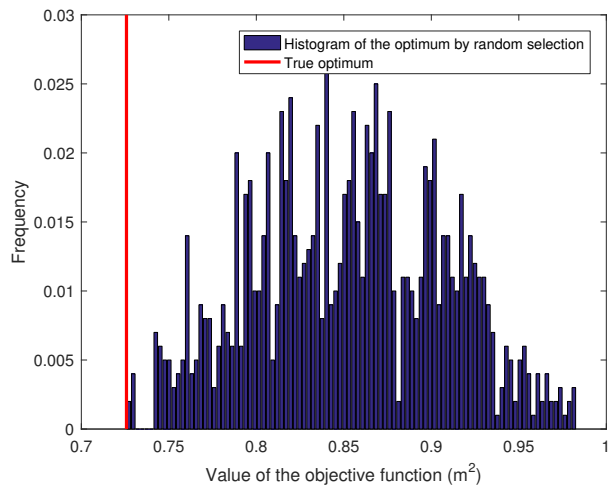}   
  \caption{Performance of random searching using equivalent computation time}
  \label{random}
\end{figure}

\indent The performance of random searching based on 1000 simulations is shown in Fig. \ref{random} for a comparison with the CE method. In each simulation, the objective function is evaluated on $N_r$ randomly selected combinations of $M$ vehicles, and the optimal value is the minimum among these $N_r$ values. The computation time and performance of this random searching depends on the fixed parameter $N_r$. A large $N_r$ would take long computation time but good performance. For a fair comparison with the CE method, this parameter is determined such that the computation time is approximately equal to that required by the CE method, which is about 5000. Fig. \ref{random} shows that only a very small percent of the optimal values obtained through the random searching are close to the true optimal value, which reflects the good performance of the CE method.

\subsection{Comparison between the two optimization algorithms}
In the previous sections, two optimization algorithms have been presented. Their scopes of application and performances are summarized here.\\
\indent The B\&B method is applied to the case where all the vehicles have the same non-common error variance. Although it is rare, in practice, to have exactly same non-common error variance for all the connected vehicles, the B\&B method is expected to find a near optimal solution as long as the variances are approximately the same. The B\&B method is a deterministic searching algorithm, which is guaranteed to find the optimal solution as long as the assumptions are satisfied. In the worst case, it would be necessary to search the solution space exhaustively, which would result in exponential computational complexity. In practice, however, the B\&B method takes much less time than the brute-force searching does.\\ 
\indent In contrast, the CE method with the pre-selection can also be applied when the vehicles have different non-common error variances. It is a stochastic algorithm, which does not guarantee an optimal solution. In practice, however, it finds solutions of high quality. The computational complexity is fixed once the parameters of the problem are given.\\
\indent For vehicle selection problems of practical scales, the average computational time of both these two algorithms is of $O(1)$ s. Thus they are promising for real-time application.
\section{CONCLUSIONS}
In this paper, a theoretical framework for evaluating and optimizing the effect of road constraints on the CMM localization accuracy is established. The major contributions and findings of this work are summarized:
\begin{enumerate}
 \item A closed-form expression that expresses the mean square localization error in terms of the road angles and non-common error is derived based on a simple CMM rule, which serves as the foundation of the vehicle optimal selection problem. Based on this expression, it is proved that the optimal distribution of road angles that minimizes the localization error is the uniform distribution.
 \item A B\&B algorithm and a CE algorithm are developed to select the optimal group of vehicles that minimizes the localization error. The B\&B algorithm can efficiently find the optimal group when all the vehicles have the same non-common error variance, while the CE algorithm can efficiently find a near optimal group when the vehicles have different non-common error variances.  
\end{enumerate}




\section*{Acknowledgment}
This work is funded by the Mobility Transformation Center at the University of Michigan under grant number N021548.

\bibliography{ITSC}
\bibliographystyle{IEEEtran}

\end{document}